# The Two Faces of American Power:
# Military and Political Communication during the Cuban Missile Crisis
*Kybernetes* (forthcoming)

Michaël Deinema & Loet Leydesdorff
Amsterdam School of Communications Research (ASCoR), University of Amsterdam
Kloveniersburgwal 48, 1012 CX  Amsterdam, The Netherlands

**Structured abstract; research paper**

**Purpose**
The mismatches between political discourse and military momentum in the American handling of the Cuban missile crisis are explained by using the model of the potential autopoiesis of subsystems. Under wartime conditions, the codes of political and military communications can increasingly be differentiated.

**Design/methodology/approach**
The model of a further differentiation between political and military power is developed on the basis of a detailed description of the Cuban missile crisis. We introduce the concept of a "semi-dormant autopoiesis" for the difference in the dynamics between peacetime and wartime conditions.

**Findings**
Several dangerous incidents during the crisis can be explained by a sociocybernetic model focusing on communication and control, but not by using an organization-theoretical approach. The further differentiation of the military as a subsystem became possible in the course of the twentieth century because of ongoing learning processes about previous wars.

**Practical implications**
Politicians should not underestimate autonomous military processes or the significance of standing orders. In order to continually produce communications within the military, communication partners are needed that stand outside of the hierarchy, and this role can be fulfilled by an enemy. A reflexively imagined enemy can thus reinforce the autopoiesis of the military subsystem.

**Originality/value**
The paper shows that civilian control over military affairs has become structurally problematic and offers a sociocybernetic explanation of the missile crisis. The potential alternation in the dynamics under peacetime and wartime conditions brings historical specificity back on the agenda of social systems theory.

**Introduction**

The development of military technologies and the increased capability of states to mobilize young men for their armies sparked an unparalleled military race after the Franco-Prussian war of 1870 and the subsequent unification of Germany. This arms race went hand in hand with a strong rationalization of the army. During the period leading up to the First World War, western countries developed chiefs of staff, rigid chains of command, secret codes, and so on. During the Second World War these organizational innovations were further elaborated, reaching a culmination during the Cold War. However rational such developments may seem from a military perspective—bearing in mind the degree of international competition and the mutual threat—they increasingly reconstructed the military into a system virtually black-boxed for political control, with its own language, authority, regulations, and secrets. Because of the military's enormous powers of destruction, such independence can have drastic consequences. President Eisenhower has become well known for his warnings against uncontrollable military pressures in the aftermath of the Sputnik shock of 1957 (York, 1970).

The danger of a mismatch between military momentum and political control was painfully demonstrated during the Cuban missile crisis, when the U.S. military heads were operating with an agenda different from that of the Kennedy administration, and generally seem to have misunderstood Kennedy's intentions. In this study, we examine the difficulties that the administration had in communicating its intentions and exercising effective control over the U.S. military forces. These problems and misunderstandings almost propelled the world inadvertently into World War III. We argue that these difficulties can be attributed to systemic differences between military and political modes of communication. By using Niklas Luhmann's sociological theory of communication, American politics and its military will be analyzed as *systems* of communication with different codes (Luhmann, 1984).

The aims of this study are twofold. An attempt is made to contribute to sociocybernetic theory by examining, in a post-World War II situation, the differences and relations between politics and military operations. The tensions between these two modes of operation within the power-structure have been largely ignored in literature of social systems theory because insufficient distinction has been made between peacetime and wartime conditions. By analyzing the dangerous behavior of the U.S. government and military during the Cuban missile crisis we wish secondly to show that a systems-theoretical perspective provides more explanatory power than previous accounts of this same case. At present, the authoritative account of the Cuban missile crisis is Graham T. Allison's (1971) *The Essence of Decision.* In this study Allison employed three separate models of analysis or paradigms in order to assess the strengths and weaknesses of these models for political analysis and to shed light on different aspects of the Cuban missile crisis. Of these three models the second, the organizational model, is most significant to our analysis as it provided Allison's most compelling explanation of the mishaps in the behavior of the U.S. government.[1]

Allison's organizational model viewed government action not as the direct result of the decisions of government leaders, but rather as the outputs of the large organizations that make up the state, such as the Navy, the C.I.A., and the State Department. These organizations are "activated" by the decisions of government leaders but otherwise operate according to pre-established plans and operating rules that may not be appropriate in every situation. This configuration explains the occurrence of many actions that were not intended by the political leadership. Equally important, these organizations act as the eyes and ears of the government leaders, providing them with information on the problems at hand and defining the alternatives open to them, and thereby act as constraints on the rationality of government action and the decisions of government leaders.

However convincing this analysis seems, on closer inspection it includes several contradictions. These problems are due to the fact that, unlike Luhmann's sociology, the organizational model is not designed to describe interaction in different social contexts—which is so important to the case at hand—but rather to improve "rational" management. The most important of these problems involves the dynamic of political decision making. According to the organizational model the decisions of government leaders are mere derivatives of organizationally defined alternatives and can therefore be formulated only as *general* instructions. From Allison's own study, however, it becomes very clear that the Kennedy administration did have *very specific* wishes and demands that differed significantly from the proposals of the army organizations. The dynamic which produced these wishes and demands will be examined here. Furthermore, we will show that there are no indications of differences in perspectives among the Navy, the Ground Forces, and the Air Force officials as one would have expected on the basis of Allison's model. On these grounds and others that will become apparent in the course of the discussion below, we propose that the potentially fatal incidents of the Cuban missile crisis can more fully be explained within a sociocybernetic framework, that is, as the result of structural differences between political and military communications.

**Two Shapes of Power**

In the way the term is conventionally understood, the possession and processing of "power" is undoubtedly the main defining element of both politics and the military. In the case of the Cuban missile crisis this power presented itself most vividly in the ability of the American and the Soviet leadership to order the deaths of many millions of people, thereby effectively ordering the destruction of their adversary's society. Both militaries would have been able to execute such an Armageddon in a matter of hours, if not less. Although potentially present, this most extreme form of power use has fortunately never been actualized on this scale in the history of the world.

The conventional understanding of "power" is not identical to the way in which Luhmann (1990a, 2000) employs the term. In his sociological analysis power is considered as a symbolically generalized medium of inter-human communications (Parsons, 1963a), and the political system as the specialized social system in which power is processed and "legitimately" expressed. According to Luhmann (1990a: 167) the core function of politics is to make "collectively binding decisions." Power must therefore be seen as the tool that mediates between the different interests or opinions that compete to form these collectively binding decisions. Perhaps mindful of the analysis of the famous German military theorist Carl von Clausewitz (1832), who proposed to see military force as simply an extension of politics, scholars using this perspective have not felt the need to make a distinction between political and military operations, or to consider the potential difference in these relations under wartime and peacetime conditions.

We propose that Von Clausewitz's subjection of military action and war to political objectives is not a timeless constant. We intend to show that the military's hierarchical dependence on politics gave way to a functional independence under relatively recent historical conditions. This independence signaled the emergence of a "power"-processing subsystem of society able to compete with politics. As an analogous development, we point to the gradual separation of science from religion as a "truth"-processing system (Luhmann, 1990b) and the serious competition between the two that has resulted from this separation, e.g., the conflicts between the theory of evolution and creationism.[2]

If military operations presently no longer would conform to the creation of collectively binding decision making and can under certain conditions operate independently of the political system, we have to offer a differentiated functionalist definition of power and consider this differentiated power as the determinant of the functions of two societal subsystems. Parsons (1963a) defined power in social systems as one among the symbolically generalized media of communication. From the perspective of society as a whole, power allows society to continually make selections out of a host of self-referentially produced alternatives for its future course (Luhmann, 2000). For those who employ it, and for every instance in which it is employed, power functions in a relative sense because other options remain available. Power comes into play when there is some form of resistance or challenge, that is, when alternatives to the existing order present themselves. In an environment of competing power-wielding units (i.e., an organization such as a state, a movement, an individual) supporting certain alternatives, the power of such a unit can then be defined as its ability to curb or subdue resistance, or in other words: the *ability to diminish, prevent or destroy* communications enacting competing alternatives in the unit's environment.[3]

During the Cold War, the United States and the Soviet Union could be considered as competing units representing alternatives for organizing society at the generalized level. The power of these two systems lied in their relative abilities to diminish or destroy each other's type of social coordination. Use of such power could be directed at the producers of the other unit's communications (i.e., people and institutions), the media through which potential communications run (e.g., telephone lines, train tracks), and the carriers of previous messages that can serve to perpetuate the specific mode of communication (i.e., codes of law, textbooks, literature, culturally significant buildings or statues, etc.). Specific attention would be directed at opposing the other unit's specialized power-wielding functions.

Whereas politics and the military both wield power as defined above, their functions can become differently defined. Power as a medium can be split into two forms. The first is the ability to diminish or destroy other units while taking over constituent elements or the media of communications of other units in order to have these other units perform and participate in accordance with the dominant alternative, which comes down to collectively binding decision making. The second is the ability to diminish or destroy other units without enlarging one's own mode of communication, i.e., by annihilating constituent elements or media of the other unit. This produces no collectively binding decisions as it destroys that which would be part of the collective. Instead, it is designed to produce a *tabula rasa* of death or complete submission. Whereas political discourse is designed to produce active participation in alternatives and partnership, the military aims to produce silent passivity in the enemy.

In the American and Soviet context at the time of the Cuban missile crisis, this distinction between hegemony and annihilation had become sharp enough for these two forms of power to be wielded by two different subsystems of society. The first form had as its institutional custodian the political subsystem of society and the military subsystem had been specialized to organize the second form of power. One should realize that in this model politics and the military, when developed beyond the incipient state of a means-ends relation, are only *sub*systems in the sense that the sum total of their operations continues to perform a common function towards society as a whole under the condition of peace. Under the extreme pressure of war or a serious threat of war, however, they have the potential to operate as fully developed subsystems, i.e., as self-organizing systems of communication. While they normally strive to maintain and organize their own modes of communication and not necessarily to perform a function to the exclusion of each other's mode of discourse, the two subsystems are organized in institutional arrangements through which they compete in contributing to the function of organizing power in society under wartime conditions.

The potentially ongoing processes of differentiation imply that relations between the subsystems can change, turn sour, and even degenerate into a power struggle at a subsocietal level. Both these operations rely heavily on the societal framework in which they are embedded and on the functions they perform in that society for their maintenance and reproduction. The differentiation between politics and the military, however, can produce tensions that are *dys*functional for society at large. The systemic insistence on self-maintenance can become particularly *counterproductive* or even dangerous when society's needs change. The two competing subsystems can be expected to provide these changes with different interpretations.

**The Cuban Missile Crisis: a brief account**

When John F. Kennedy was elected President of the United States in November 1960 the Cold War seemed to enter a lukewarm stage. After numerous conflicts during the fifties, among them the Korean War, the two superpowers were more prepared to settle their differences through diplomatic channels. However, a few thorny issues remained on both sides. Behind the Iron Curtain, West Berlin was still a capitalist stronghold, and Cuba gave the Soviet Union a footing in the Western hemisphere. In 1961, one year before the crisis, the communists erected the Berlin wall, and exiled Cubans attempted an invasion of Cuba with American support. This latter undertaking would become known as the Bay of Pigs disaster.

The blame for this fiasco provided a hard bone of contention between Kennedy's administration and the military chiefs, who would afterwards live in strong mutual distrust (Brugioni, 1990: 260). These events also made "Cuba" and "Berlin" even more politically sensitive on both sides of the ideological divide. Pressures built up inside the U.S.A. to have the administration rid the western hemisphere of the Cuban Revolution, or at least to prevent the island from becoming a forward Soviet military base. Across the Atlantic, Western European countries expressed the fear that the Americans would not risk a nuclear war to prevent a Soviet encroachment on Berlin or other areas, especially if the Soviets could launch a devastating strike from Cuba. This configuration led Kennedy to issue two public statements in September 1962 that any introduction of Soviet "offensive" weapons into Cuba would not be tolerated (White, 2001: 151-155).

After the Bay of Pigs incident, Cuba's Fidel Castro repeatedly urged Nikita Khrushchev—the Secretary-General of the Communist Party of the Soviet Union—to send weapons to Cuba that could prevent a new invasion. Khrushchev sent "defensive" weapons and crews to handle them and the United States grudgingly accepted this. At the same time, however, members of the Politburo and of the Soviet military establishment were urgently seeking means to offset the U.S.'s numerical nuclear superiority. For this reason, and to provide the Soviets with a bargaining chip for negotiations on Berlin, missiles and nuclear warheads had already been deployed to Cuba at the time Kennedy issued his warnings in September 1962 (Khrushchev, 1990: 170-174; Brugioni, 1990: 243-244).

The Soviets had hoped to present the Americans with a *fait accompli* at the end of the year, but their military build-up was discovered on October 15 when it was still in progress and when the missiles were not yet operational. When Kennedy was notified the following morning, the most dangerous fortnight of the Cold War had commenced. Kennedy was convinced that something had to be done, if not for military then for political reasons, and convened a group of his closest advisors. This group, later to be called the Executive Committee of the National Security Council (ExCom), deliberated in secrecy for a week together with the highest circles of the American military and intelligence establishments on the course of action to be taken.

Originally favoring the option of an air strike to remove the missiles by force, the deliberations in ExCom slowly shifted towards supporting the less impulsive option of instituting a naval blockade to prevent more offensive military equipment from entering Cuba. This had the advantage of sending a strong message to the Soviet Union and the rest of the world of the determination on the side of the U.S. to see these missiles removed, while at the same time leaving the Soviets with some time to deliberate and find a graceful exit from the conflict. If the build-up persisted, further steps could still be taken. The main disadvantage from the American point of view was of course the fact that the blockade, or "quarantine" as is was called, did nothing directly to remove the missiles that were already in Cuba or to prevent them from becoming operational.

Kennedy chose to pursue the "prudent" course of a naval quarantine despite the fact that a consensus in favor of this option had only been reached painfully within ExCom and that the highest military organ, the Joint Chiefs of Staff (JCS), persisted in calling for a massive air strike followed by an invasion of Cuba. He announced his decision to the American public, the Soviets, and the rest of the world on Monday October 22, one week after intelligence officials had first discerned the missile sites from photographs taken by a U-2 spy plane. At the moment this televised announcement was made, U.S. military forces all over the world moved to Defense Condition 3 (DefCon 3), only two steps away from general war. Naval forces moved into position to institute the quarantine, which was to take effect on Wednesday morning; the Air Force and infantry were readied to commence further action against Cuba; and conventional and strategic (nuclear-armed) forces were prepared worldwide for any Soviet retaliation, such as a move against Berlin.

An extremely tense week followed in which the world came very close to a massive nuclear war at several moments. With Khrushchev at first unwilling to yield to the blockade and Kennedy not prepared to make concessions given the superior American nuclear arsenal while under the close scrutiny of the world, the two superpowers were moving towards disaster. In the meantime, the two powers frantically tried to employ diplomacy bi-laterally, at the United Nations, and with Cuba and Turkey (since Turkey's American missile sites might have to be removed in order to resolve the conflict evenhandedly). In addition to confronting each other and the other allies involved, the two heads of state also had to deal with their own officials, some of whom seemed to support moves which might lead to an escalation.

Several potentially catastrophic incidents occurred which were out of control for the two political leaderships. Most of these incidents resulted from the actions and routines of the respective militaries. The realization on both sides that these incidents might not have been the result of conscious decisions from the respective highest authorities helped contain the seriousness of the situation. Another important factor in this respect was the tendency to carefully examine alternatives to each proposed action or reaction. Precisely this self-restraint in the light of so much threat, and the profound fear and sense of responsibility felt by both Kennedy and Khrushchev, paved the way for an eventual solution of the crisis. The two superpowers eventually agreed that the missiles and the nuclear weapons in Cuba would be removed in return for an American pledge that the U.S.A. would never invade Cuba and that the American missiles in Turkey would be removed within six months, under the pretext of their obsolescence. With this agreement the world as we know it was saved.

**The Political Discourse of Persuasion**

The eventual conclusion of the Cuban missile crisis may seem to warrant some apprehension about the historical seriousness of the conflict. After all, it is hard to believe how intelligent and sane people could deliberately enter into a nuclear war. The simplicity of the solution and the virtual inconceivability of the terrible alternative the world faced could with hindsight mitigate our evaluation of the difficulties involved in reaching this prudent course. The evaluative perspective of hindsight, however, would fail to do justice to the historic circumstances in which the deliberations were held and, more seriously, might make it impossible for us to gain theoretical understanding from these events. Furthermore, it would entail an analytical error in examining the actual decision-making process in a crisis on the basis of its outcome. On the American side at least, and this side will be our focus, many different perspectives and evaluations of priorities emerged, conflicted, and interacted with each other during the process, producing different outcomes at different moments during the deliberations.

To see whether or not the model of the political system proposed above is applicable to the actual ExCom deliberations, we first need to examine its implications a bit further. The form of power that we have assigned to the political system can be summarized under the broad heading of "persuasion." The political system operates at different levels. One of these is the level of international relations in which national governments vie for influence as units. Of course, it is not enough for government officials to know that they have to persuade leaders and people from other nations; they also need to know the content of what they have to persuade them. The complex whole of national interests is determined by the nation's dominant modes of interaction. Therefore the political discourse of the U.S. requires representations within itself of the other major modes of communication in this society. These representations have been institutionalized within the U.S. government as departments and committees, but are also present as culturally generalized ethical considerations and values. To formulate a "plan to persuade" in complex situations, the national units within the political system must attempt to homogenize these very different representations contained within them. They need references to their own preferred alternatives.

In order to achieve this integration, the system can be expected to create special functions. In the United States, the most prominent of these functions is that of the nationally elected President. Along these lines, a hierarchical structure is created within the national government that can be reflected in the structure of the departments, which are differentiated as well. Here the secretary of each department performs the integrative function. This hierarchical organization of the state, however, does not reflect a hierarchical organization of society as a whole.

Because of the need to keep American knowledge of the missiles in Cuba secret until specific action was decided upon, the President, whose moves were always carefully followed by the media, had to keep up his regular schedule and was unable to devote all his attention to the crisis (May and Zelikow, 2002: 72-73). This meant that another mechanism was needed to assist him in integrating the various perspectives and to provide him with a limited number of clear alternatives. Such a mechanism was furnished by the ad hoc creation of ExCom, mainly organized by the President's brother, Attorney General Robert Kennedy. In attempting to set a course of action, ExCom had to integrate military, diplomatic, legal, ethical, and other types of considerations. They continued to do so after Kennedy had announced his decision to the world to institute a quarantine. Even though the President could now direct all his attention to the crisis, he would keep ExCom for the remainder of the crisis and act as its chairman.

Virtually all ExCom meetings were conducted with the secretaries and undersecretaries of the Departments of State, Defense, Justice, and the Treasury as main participants, together with representatives of the United States' diplomatic services, intelligence community, and military. During the deliberations, different topics of a very diverse nature came up. Although most attention was directed to military and diplomatic affairs, questions also arose about whether the government had the legal authority to order an air strike without a Congressional vote, whether or not there would be demonstrations in Latin American capitals against U.S. actions, and about the importance of the opinions of the American and British media. A very important argument against the option of an *unannounced* air strike was an ethical and very American one: the United States would be acting against its own tradition if it ordered a "Pearl Harbor attack" on Cuba (Brugioni, 1990: 242).

All kinds of issues can enter the political discourse. In terms of systems theory, this ability can be called the capacity to handle a great deal of complexity. As in all systems, this capacity is achieved through further differentiation; in this case a differentiation of the state unit into departments and committees. Handling such a high level of complexity, however, also requires a form of communication that can bear the burden by reducing the complexity in a functional mode. Much political interaction occurs in the form of threats and offers, but within the same unit, the political system achieves this reduction of complexity by coding the communication in terms of *arguments*, *advice* and *recommendations* as its main forms of interaction. These formats allow the integrative mechanisms to assess and weigh very different aspects of a problem. Although external interactions can be coded in another system's form of communication, all representations within the political system operate according to this *qualitative* form. The external systems have to be represented within the political discourse even if they use other forms of codification internally. For example, the military was represented in ExCom by the chairman of the Joint Chiefs of Staff (JCS), General Maxwell Taylor, who had to argue in this context with representatives of other interests representing, for example, commercial and trade interests in keeping the peace.

Arguments can evoke many possible reactions from other participants in the network. Not only can many different types of input be processed by the political system using its specific codifications, but the political communications have multitudinous "meanings" assigned to them. Each input has a very high *meaning density* when it enters the political system; every input can evoke a large range of responses. Because of this fluid nature of political interactions, political decisions are likely to remain *provisional*, given that new input from any direction could add new considerations and that each communication can be interpreted inside the system in so many ways. Therefore, unshakeable positions are uncommon. As Robert F. Kennedy (1969: 31) would formulate it with hindsight:

> "(…) none was consistent in his opinion from the very beginning to the very end. That kind of open, unfettered mind was essential."

Even presidential decisions are prone to the fate of remaining provisional. For example, when Kennedy heard that the Soviets demanded the removal of the American missiles in Turkey, he became angry because he had ordered the removal of these missiles a year before the crisis. However, his Secretary of State, Dean Rusk, had neglected to arrange the removal due to resistance from the Turks (Brugioni, 1990: 466-467). Removing them now under Soviet threat could be perceived by the rest of the world as a sign of American weakness. At this point in history, the hierarchies within the state could not be very strict, because the state, like the political system as a whole, attempts to organize representations of a society that was progressively organizing itself less along hierarchic and more along functional lines.

The continuous fluidity and provisionality of political decision-making stands squarely opposed to the military expectation to be provided with a clear definition of objectives. However, considering political deliberations as part of a subsystem that is interested mainly in persuasion, can account for the way in which American deliberations were conducted during the Cuban missile crisis. It also accounts for the common perception of politicians as engaging in endless discussions and their tendency to fail to stand firm in action.

**Military Command and Control**

When an international conflict arises, a nation's political and military leaders are forced to work together intensively. This close cooperation does not mean, however, that the one discourse can be subsumed as a subsystem to the other as in a means-ends relation. The difference in the communicative mode is too fundamental to warrant such a conclusion. This difference has crystallized institutionally to the point that the political system needs a representation of the military system (institutionalized in Departments of Defense) and the military system needs a body to confer with the political system (i.e., the Joint Chiefs of Staff). The interface serves a translation between the political discourse that revolves around persuasion, and the military mode of communication that revolves around destruction, command, and control.

At the beginning of the Cuban missile crisis there already existed a plan for an attack on Cuba (Allison 1971: 125). In the planning of such scenarios, the reasons why the plan might be undertaken are considered irrelevant. In other words, from a *strictly* military point of view, no justification is required. This rationality is not only shown by the military's extreme belligerence during the crisis, but also by general military traits, such as the use of a morally indiscriminate terminology. Terms such as "casualties", "collateral damage", "rules of engagement" and "Oplans" are relatively devoid of value judgmental connotations.

General Curtis LeMay, the Air Force Chief of Staff at the time of the crisis, is reported to have even shouted "We lost! We ought to just go in there today and knock 'em off!" when he heard that the Soviets had eventually agreed to remove the missiles and that the crisis was over (Rhodes, 1995: 575). Although LeMay's position was extreme even among military men, it does prove a point about military interaction. The focus of the military as a whole is primarily on the destruction of the enemy's resistance and on the national defense against such destruction. It creates a host of plans for attack and defense, and executes those that it is allowed to execute by the government.

LeMay's counterbalance at the JCS was the chairman, General Taylor, who was also a permanent member of ExCom. Taylor had been appointed by President Kennedy himself after the Bay of Pigs fiasco and therefore had to keep the political objectives continuously in mind. Time and again he would remind the other chiefs of staff of the political consequences of the actions they proposed. During these deliberations he would also try to explain to them how differently military men and politicians would approach problems, usually to little avail (Brugioni, 1990: 262).

We thus argue that the military system has an orientation fundamentally different from that of the political system and that this military orientation revolves around destruction. When competing power-wielding units are considered as systems of interaction, this destruction is not necessarily directed at people or physical objects. A blockade, for example, destroys the adversary's lanes of exchange and is therefore a military action, even if no physical destruction takes place. It can be compared with recent developments in cyber-warfare plans aimed at destroying enemy communication.

From this premise about the rationality prevailing in military communications, several implications follow. First, attempts at destruction are bound to meet with fiercer resistance from the adversary than attempts at persuasion. Persuasion aims at some agreement by the subjects to be persuaded, whereas destruction is inherently perceived as aggressive. This makes military action *dangerous* and increases the stakes for all involved. Secondly, destruction is a relatively *straightforward* process; fewer factors are involved than in the case of persuasion, which can be seen as a process of *de*construction and *re*construction of legitimacy. Like a political organization, the military requires forms of autoreference and reference to its competitor. But unlike a political organization, it does not require references to its own preferred alternatives. It is not representative. Consequently, the military system can afford to handle *less* complexity than the political system. These two factors are decisive for the main forms of internal interaction employed by organizations within this system, which are *plans, regulations*, and *orders*.

The restricted types of meaningful input enable the military to devise long-term plans (strategy) and work out their plans in great detail and with more stability. Not so many factors can significantly disturb military operations. This closure of military interactions makes regulations, which are preserved complexes of previous orders, useful because situations will often occur within the same configuration of a limited number of meaningful factors. Plans and regulations are not only possible, but also necessary in the military system because of the inherent danger in military operations. If actions are not well coordinated, more soldiers may die and consequently the morale may drop. Both these developments are detrimental to the system as a whole. To achieve good coordination, plans and regulations are needed and orders must be followed.

If military coordination is to succeed, orders need to be distinguishable from other forms of communication. The distinguishing feature of an order is that the rank of the originator is superior to that of the recipient, and this automatically makes a strictly hierarchical structure, a chain of command, necessary to any military organization. An order is thus a communication with a rank attached to it, so that it only allows for a very limited range of responses. It naturally excludes the exchange of arguments.

The immediacy of situations facing soldiers usually leaves little time for deliberation anyway and makes them even more inclined to follow orders. This need for *speed* is a theme that runs through virtually all military communications. It increases the utility of orders and officers' knowledge of plans and rules of engagement, so that no time is wasted on detailed explanation. Modern technology has made it possible to devastate the largest nations in the world, including their war-making capabilities, overnight (Rhodes, 1995:

575). It was therefore understandable that during the crisis the JCS emphasized the importance of speed and the specific military forms of communication so strongly.

Orders have a clear origin and destination. Their reproduction is as finite as the chain of command, except when the execution of orders produces results of specific meaning to the military rationale. When no such meaningful results are available the production of military communication depends on original decisions produced externally, the prime movers being politicians, and the military "system" is not independent or self-organizing, but an extension of politics. This seems to be the case in times of peace. The operations of the military in peacetime are the result of politically approved programs that do not by themselves produce new programs and can offer continuity only in the form of repetition. In order to continually produce communications within the military, communications partners are needed that stand outside of a unilinear hierarchy, and this role can be fulfilled by an enemy or a virtual enemy.

The confrontation between enemies can thus contribute to bringing the military system into a self-organizing mode. Meaningful "results" occur most clearly during actual military confrontations. The elementary units of military confrontations between opposing military forces are *hostilities*, which are similar to orders in being expressions of the medium of silencing/destructive power or *force*. Hostilities are produced by orders, which are recursively produced by hostilities, as pre-established plans, regulations and objectives refer to hostilities and assign military meaning to them. Reflexively, the confrontation of orders, plans and regulations with actual hostilities produces new meanings for potential future hostilities and orders.

During a war (a state in which hostility is sanctioned) this could lead without interference to an endless and *independent* chain of military interactions if plans, regulations and objectives are adequately developed to give meaning to what happens "on the ground." These meanings are then not produced by non-military interference or by the inspiration of individual commanders, but are the results of the systems own inner logic. The confrontation between two opposing military organizations can thus form a self-organizing or autopoietic system. The enemies are an integral part of this autopoietic system, which exists until one side is destroyed or peace is made. The importance of pre-established regulations, which are basically complexes of preserved previous orders, in propelling the military system into autopoietic mode implies an historical development from an extension of politics to an autopoietic system, with a stage of semi-dormant autopoiesis (autopoiesis only in wartime) as the missing link.

**The Relations Between the Military and Political Subsystems**

The most obvious aspect of the interaction *between* the political and military systems is the constitutional authority of the political system. The explanation for the prevalence of the political system finds its origin in the larger social framework in which both these systems are embedded. The political system is more closely attached to the rest of society than the military because of its internal need for representations of the other systems of society. The process of homogenizing the concerns of these different systems, as perceived by the political system at least, or of somehow fitting them all into a larger picture in order to select or export a representation of the society, makes the political system the natural forum for decision-making within society under peacetime conditions.

The political system not only generates representations of society, but also adjusts society itself to conform to this representation (Scott, 1998: 80-83). The political system can thus be considered as the center of its society's network; its influence within society arises from the use that all other systems are able to make of it for their own ends (Parsons, 1963b). It forms a natural link for society's systems to call upon when they require the allocation of resources from another system. The political representation of the military system, for example, which can be considered as an interactive interface, functions as the military's tie to the market and other social subsystems. When the military needs money, generally healthier recruits or more highly educated physicists to design new weaponry, it appeals to the political system to arrange this for them. The quid pro quo for this service is recognition of the ultimate authority of politics over military affairs, which in turn is very useful to the political system's own goals.

The relation of the political system with the military can, under normal circumstances, be expected to be twofold. First, the political system entertains representations of the importance of the military's position in the political evaluation of society. This assigned importance then decides the degree to which political units can attempt to export that military system or expand its military influence. For example, the German army can claim only a very modest role in the political configuration of the German Federal Republic, whereas the United States is able to control and organize NATO's forces in accordance with its own model to such an extent that the American President appoints the Supreme Allied Commander.

This strategic position of the U.S. played a role during the Cuban missile crisis. The commander of NATO at that time was Lauris Norstad, who had been appointed by President Eisenhower. President Kennedy personally delayed the scheduled appointment of General Lemnitzer, the ex-chairman of the JCS, as Norstad's successor on the day before the quarantine of Cuba went into effect (May and Zelikow, 2002: 216). At that moment, a further strengthening of the integration of the international arena with the American military was no longer in the interests of American politics.

Secondly, political units use their interfaces with and their authority over the military to more effectively persuade other political actors. The threat of authorizing military action is a powerful tool to convince others to adopt foreign customs, such as free trade or democracy. The military system can also be used to fight competing units over territory or resources that a political unit wants to incorporate, or to deter attacks. In the spirit of Von Clausewitz, the *exercise* of military power is for the political system always a means to an end (Von Clausewitz, 1832). The expansion, however, of the military *model* is, as an expression of hegemonic power, an end in itself.

As explained above, the continuity of military interactions was long dependent on political decisions. It is not difficult to sketch the historical development that weakened and eventually broke this dependence in the industrialized world. The rationalization of military affairs after 1870 led to the detailed specification of grand strategies (e.g., the *Schlieffen Plan* of 1905) and the establishment of routines for large-scale mobilizations. In terms of actual combat, however, the stand-off in the trenches of World War I painfully demonstrated the limitations of fighting according to the principles of previous wars. Political action was still required to prevent the endless repetition of doomed assaults.

During the Interbellum armies were refashioned to be more responsive to technological developments and more adaptable. The extreme ravages of World War II and the creation of nuclear weapons subsequently made the costs of total war seem prohibitively high to politicians. As a result, the definition of "war" was broadened to encompass *potentially* hostile behavior. Potential hostilities became as militarily meaningful as actual hostilities. The Cold War was invented, thereby adding a reflexive level to military operations, the level of the *virtual enemy*, whose behavior in an imagined wartime situation had to be constantly imagined and reimagined on the basis of military intelligence. As Eisenhower had predicted, the new situation developed into a relentless arms race after the Sputnick shock. This arms race effectively made the military autopoietic under the conditions of formal peace.

Over roughly the same period, the political system made progressively less use of hierarchical forms of organizations and interaction (at least in the West). The transition from constitutionally self-limiting "nightwatch states" to socially involved welfare states occurred (Luhmann 1990a: 169-171). Because the political system—still mainly organized in national units—could not fully control the international developments of social systems it was increasingly concerned with, the political system needed to function less hierarchically. This development widened the schism with the military system.

But eEven when the military system achieves a state of autopoiesis and therefore becomes a functionally differentiated system—perhaps, only temporarily—it remains structurally coupled to the political system in the longer run. Being a consumer rather than a producer, the military system needs the political system to secure its provisions. But like any inferior partner in a relationship, especially one with so much inherent power, the military can turn on its master or at least try to break free. This could happen if it has repeatedly been frustrated by the political system, which was in fact the case in the period leading up to the Cuban missile crisis. When the Bay of Pigs invasion was being planned, a significant window seemed to open up for U.S. military deployment. Fighter-planes were already manned to assist the Cuban rebels when the President decided against *overt* military support (White, 2001: 29). The military machine had started up, but was ordered suddenly to halt. It had accomplished nothing. Nevertheless, it received part of the blame for the fiasco. Another instance was the prospect of military intervention in Laos, which was also blocked by the administration (May and Zelikow, 2002: 122).

Given these experiences, the JCS did not want to be restrained again when the missile crisis started, and the chiefs of staff were prepared to risk a global war for the chance to attack Cuba. Consequently, they pressured the President to commit to his own statements of the previous September, that he would not allow offensive weapons in Cuba, and were overwhelmingly convinced that an expanded air strike was the only appropriate action. Except for General Taylor, they also insisted on a follow-up invasion of Cuba. Even Navy Chief of Staff Admiral George Anderson, who might have been expected to support the blockade option as a chance to show off the navy's competence, tried to convince Kennedy that an expanded air strike was a preferable alternative (May and Zelikow, 2002: 114-115).

**Two American Faces of Conflict**

> "An invasion would have been a mistake. But the military are mad. They wanted to do this." - President John F. Kennedy (Allison, 1969: 364)

This statement by Kennedy after the crisis illustrates his assessment of the advice that he received from the Joint Chiefs of Staff. His brother was even more forceful in his rejection, asserting in his memoirs of the crisis that the military representatives to the President, apart from General Taylor, always assumed that the Russians and the Cubans would not respond to attacks and that war was in the American national interest. In another chapter he recalled all the times "the military [would] take positions which, if wrong, had the advantage that no one would be around at the end to know" (Kennedy, 1969: 48 and 119).

Not all disagreements and misunderstandings came purely from the JCS's desire to go to war. The military emphasis on speed, their rules and regulations, and the significance of orders conflicted sharply with political provisionality and the tendency to deliberate. When President Kennedy tried to make sure during the crisis that the American nuclear missiles in Turkey and Italy could not be fired without his express orders, the JCS protested heavily because that implied to them a lack of confidence in standing orders (May and Zelikow, 2002, 141). When both Kennedys objected to performing a "Pearl Harbor attack" on the missile sites and on Cuban airfields, General Taylor argued that secrecy was an *integral* part of military operations and that giving up this advantage was simply stupid. The President, however, "put the JCS cynicism in his pocket and looked for rainbows." (Brugioni, 1990: 245).

Differences in thinking were perhaps most clearly noted by former President Dwight D. Eisenhower, who supported the proposed blockade even though as a former general he added that he thought a surprise attack would be the best thing to do *militarily* (May and Zelikow, 2002: 142). Above all, the military pressed for a *quick* decision from the President, fearing that Cuban and Soviet preparations were steadily making an attack more difficult and that the missiles might be hidden from sight overnight, while the President's intention was to give the Soviets some time to retreat.

Another point of contention was the actual significance of nuclear missiles in Cuba. Whereas the military considered them a considerable asset to the overall Soviet threat, the politicians seemed not too concerned about the striking power of these missiles. In the minds of the President, Secretary of State Dean Rusk, Defense Secretary Robert McNamara, and others, the missiles in Cuba could in case of nuclear war only add to an already *unacceptable* blow deliverable by missiles based within Russia. But this consideration did not prevent the politicians from taking the diplomatic and electoral significance of the situation extremely seriously (May and Zelikow, 2002: 60-62).

The historical examples invoked by the members of the two subsystems were also very different. To his trusted advisors Kennedy, on the one side, mentioned several times during the crisis that his reading of Barbara Tuchman's (1963) book *The Guns of August*, about the outbreak of the First World War, made him very fearful of accidentally initiating a series of events that would lead to general war. His concern not to appease the Soviets, as the British and the French governments had done with Hitler in 1938, also arose from fear of having to face an even greater conflict later on if firm action was not undertaken at this time. General LeMay, on the other side, appealed exclusively to the lessons of appeasement, specifically recalling the President's father's personal role in this historical mistake, but only in order to make another point: notably that war was unavoidable. For the same reason, the chiefs looked back at the more recent events of the Bay of Pigs and Laos as their points of reference. The relevant lesson for them was not how war could be prevented, but rather how it could be started.

The belligerence of the military reflects both an assertivity perhaps brought about by its newfound partial independence as well as a somewhat outdated desire for war in order to gain virtually complete independence of operation. The fact that Admiral Anderson shared the opinions of the other JCS members that the air strike was the preferable alternative negates the claim of Allison's organizational model that the Navy (and by extension its chief commander) was motivated mainly by a sense of competition with other defense organizations. It seems that in this case at least such a sense was trumped by a generalized military rationale.

**The Dangers of Misunderstanding**

True danger of escalation came from incidents in the field that could easily be misunderstood by the Soviets. These incidents were the result of military operating rules and the political ignorance or neglect thereof. One of these operating rules stated that a naval blockade line should be out of reach of enemy aircraft, which amounted to an arc 500 miles out from Cuba (Brugioni, 1990: 381). When the British ambassador Ormsby-Gore suggested a day before the quarantine went into effect that the arc should be drawn closer to Cuba in order to give Khrushchev more time to decide what to do, the U.S. President agreed. Admiral Anderson, the Navy Chief of Staff, angrily protested against this transgression of naval guidelines, before yielding to Kennedy's wishes. Allison has even suggested that the Navy did not move some of its intercepting ships at all, increasing the chances of a confrontation (Allison, 1971:130). Naval ships trailed Soviet submarines far beyond even the 500-mile arc. Upon hearing this information, it was Defense Secretary McNamara's turn to become furious. He called it an irresponsible and unordered provocation. Admiral Anderson replied that it was standard naval procedure in case of a blockade. McNamara did not accept this as a valid argument (Brugioni, 1990: 415-417).

In order to achieve its purpose of launching war, even standard regulations were disobeyed by the military. On Wednesday, October 24, General Thomas Powers, commander of the Strategic Air Command, issued the order to his troops to move up to DefCon 2, the last preparatory step before general war. Powers' order, however, was transmitted *in the clear*, in such a way that the Soviets monitoring American transmissions could understand it. This went against standard procedure that dictates that a rise in Defense Condition should be transmitted *in code*. This insolent action on the part of Powers can only be interpreted as a deliberate provocation, since the prospect of the American nuclear fleet in full readiness could have made the Soviets decide that an escalation on their part had become unavoidable (Blight, Allyn, and Welch, 1993: 468; Rhodes, 1995: 572-573).

The political leadership acted irresponsibly as well in the sense that they were often ambiguous in their instructions to the military. They did not fully realize the significance of *standing orders*, treating them as if they had the same provisional status as political communications. In a briefing of ExCom before the quarantine went into effect, Admiral Anderson related the Navy's intention to fire on ships in order to disable their rudder if they tried to proceed past the quarantine line. The President agreed and issued the appropriate orders. When the situation actually occurred at the height of the crisis and Anderson prepared to give the order to fire, the strongest confrontation between Anderson and McNamara took place. Anderson had had it with McNamara and told the Secretary that he should not interfere, as the Navy had been handling blockades "since John Paul Jones". McNamara replied that he did not "give a damn about John Paul Jones" and that there would be no firing on Soviet ships (May and Zelikow, 2002: 140 and Brugioni, 1990: 474).

On the same day an American spy plane was shot down over Cuba and the pilot killed. The JCS had made it very clear during an earlier meeting that if their planes were shot at, an immediate retaliatory strike would follow on the Surface-to-Air Missile (SAM) site from which shots were fired. When General LeMay was informed about the incident, he immediately readied fighter planes to destroy the SAM site. Just before these planes took off, a frantic call came from the White House instructing him to ignore the standing order and *not* to proceed with the attack. Understandably from their perspective, LeMay and Taylor were now furious (Brugioni, 1990: 463-464).

The most dangerous incident during the crisis was the result of a standing order being executed. Three days into the quarantine and with tensions heightened, the SAC allowed a test launch of an Intercontinental Ballistic Missile (ICBM) across the Pacific Ocean. At this point in the crisis, the Vandenberg base, from which this missile was launched, was already fully armed for an attack on the Soviet Union. The Soviets therefore could not be sure that this missile was not directed at them and carrying a nuclear warhead. Despite the obvious danger of such an action at this moment in time, the Air Force proceeded according to their missile-testing schedule, which had been formally approved by the President long *before* the beginning of the crisis.

**Conclusions and discussion**

At a conference with Cuban and Russian officials, thirty years after the crisis, Robert McNamara surely exaggerated the control exercised by the civilian government over the military when he claimed that:

> "There's nothing—and I mean literally nothing—involving any significant military action that did not occur as a result of the decision of the President, through the Secretary of Defense." (Blight, Allyn and Welch, 1993: 158)

Perhaps in retrospect a missile test or General Powers' unsanctioned reporting of a rise in alert may not seem like significant military actions until we remember that, if these had sparked off a nuclear war, there might have been nobody left to take a retrospective perspective. The underestimation of the differences in focus between the military and political systems posed serious danger during the missile crisis. This source of danger added to an already extremely dangerous international conflict.

Politicians should realize reflexively what they are dealing with. The problem of miscommunication between the two systems is a structural one and should be recognized as such. It cannot be solved simply by replacing military commanders whose attitude has been exceptionally dangerous. General Taylor was as much of an exception as General LeMay. Stripping the military of all its autonomy is no solution either. As much as the military attitude worried him at times, even Kennedy recognized that it might be more dangerous if the commanders would not be willing to go to war at all (Kennedy, 1969: 119). Many of the incidents described above, however, could have been prevented if the politicians had shown more caution in their handling of the military.

For the military, an order is an order, and it lacks the provisionality to which politicians are accustomed. This obligates politicians to think very carefully about their own intentions before issuing military orders, since they may not always be reversible, and to keep a careful track of standing orders. It also obligates them to study military plans, regulations and rules of engagement closely to know exactly what a specific order implies. Paradoxically, McNamara was right by claiming political leadership: most of the dangerous actions the military took during the crisis were indeed implied in the Presidential orders. Neither the President nor the Secretary, however, was aware of the implications of previously issued orders.

The expectation of a structural solution to this problem, however, seems utopian. The self-reform required to sensitize the political system to military rigidity would inevitably lead to a systemic paradox. This problem emerges from the general paradox of self-reference (Luhmann 1990a). Only through the use of arguments, advice and recommendations can the political system initiate and realize self-reform. Unfortunately, in this case, it is exactly the provisionality inherent in these forms of communication that would be in need of reform.

The same reasoning goes for military reform. The only way it would be able to lessen the rigidities of orders is through issuing orders to that effect. Even if the addition of more reflexive levels to the military system may make it less hierarchical in the future, it is unlikely it will in this way bridge the gap with a political system that remains in constant development itself, especially because the military medium (force) is inherently more hierarchical. However, due to their structural coupling the two systems will continuously adapt to each other as long as there is no upsetting international confrontation. This balance between the subsystems thus can be different under wartime or peacetime conditions.

Most of the problems facing the Americans when handling the Cuban missile crisis can be attributed to the differences in the orientation and the operations of the political and military systems. The extent of the functional differentiation of a discourse of persuasion and a process of destruction turned decidedly dysfunctional at this critical moment. This explanation has considerable advantages over earlier explanations.

Contrary to earlier theoretical models that have been applied to this crisis, our explanation enables us to view the deliberations of the government and the operations of the military within the same framework. The special characteristics of *both* systems have been examined, whereas the organizational model viewed the Kennedy administration as a "rational" entity whose rationality was however severely bounded by government organizations that presented it with a very limited number of alternatives. We have shown that the Kennedy administration in fact came up with, and eventually—although imperfectly and with much difficulty—succeeded in implementing, resolutions that deviated from any of the alternatives presented to it, for example by conceiving of a completely original type of blockade. The sociocybernetic analysis has the advantage of being specifically designed to meet the challenges of symbolic mediation, around which this case obviously revolved. Allison's (1971) organizational model instead used a static theory interested mainly in identifying organizational structures. (The two other models Allison proposed did not take a structural perspective at all.)

Our analysis is also able to resolve several contradictions in Allison's organizational analysis, which were the result of the rigidity of his model. It explains why the Commander in Chief of Naval Operations initially favored an expanded air strike followed up by an invasion, which offered far less scope for naval participation than a blockade. It also clarifies how, prior to the crisis, the Secretary of State Dean Rusk and later his undersecretary George Ball could shelve Kennedy's strict instructions to remove American missiles from Turkish soil after the Turkish Foreign Minister and the Turkish Ambassador warned them not to. Without further specification, Allison presented this incident in defense of the organizational model. It seems clear, however, that not the operational procedures of the State Department (which cannot be designed to obey foreign leaders in conflict with presidential instructions), but rather the ever shifting power differentials acting upon political discourse and the resulting provisionality of political decisions explain this failure of compliance. Finally, our account allows the reader to place the occurrences within a context of historical development, and can therefore take more information into account than the more static organizational model.

The application of the communication systems model was fruitful in this study, but our story has remained a single case study. In order to understand the larger applicability of this definition of the military and of politics in systems theoretical terms and of systems theory in general, much more work needs to be done. After examining the American side of the crisis, an interesting place to start would be to examine what happened on the Soviet and Cuban sides, and to see if their crisis handling was similarly affected by systemic problems. Such an inquiry would go a long way towards determining the real explanatory power of this model for modern conflicts and shed more light on the differentiations this analysis has proposed recursively to Luhmann's systems theory. By proposing the concept of a "semi-dormant autopoietic system" that is located clearly in time and international

"space", it reestablishes a need to differentiate sociocybernetic analysis with reference to regional contingencies and national history.

**Acknowledgement**

The authors are grateful to Marjolein 't Hart for comments on a previous version of this paper.

**Notes**

1. Allison's (1971) two other models are (1) the rational actor model that serves to highlight the strategic significance and justification of government decisions and (2) the bureaucratic model that analyzes the major policy decisions as the results of bargaining between several important players and different interests within the government.
2. Luhmann sees a fundamental incompatibility between functionalist sociology and religion, as recognition of the functions of belief destroys belief itself (Luhmann 1990c: 157). It is however recognized that modern science developed historically under the aegis of an unshakable belief in God and that scientific research was for a long time justified in explicitly religious terms (Luhmann, 1990b; Leydesdorff, 2001).
3. Other definitions of power are certainly possible. Stephen Lukes (1974), for example, exposed what he called the third dimension of power, which is radically different from our "tug-of-war" conception of power. However, it can be assumed that forms of power, hidden in an *already* constructed social reality, could not have been the basis for the construction of an explicit power-wielding discourse.